WallStreetBets: Positions or Ban

Christian J. Boylston, Beatriz E. Palacios, Plamen T. Tassev

Georgia Institute of Technology



Abstract

r/wallstreetbets (WallStreetBets or WSB) is a subreddit devoted to irreverent memes and high-risk options tradings. As of March 30, 2020 the subreddit boasts a usership of nearly 1.1 million subscribers and self-describes as "if 4chan found a Bloomberg terminal." This paper will utilize Amy Jo Kim's community design principles along with social psychology theory as frameworks to understand how this chaotic, oftentimes offensive community has developed one of the largest and most loyal user bases on the platform. We further will argue that humor plays a vital role in promoting in-group cohesion and in providing an unconventional third place for traders (and thinly veiled gamblers) to seek support from each other in the form of vulgar, yet good-humored taunting.

WallStreetBets: Positions or Ban

r/wallstreetbets (WallStreetBets or WSB) is a subreddit devoted to irreverent memes and high-risk options tradings. As of March 30, 2020 the subreddit boasts a usership of nearly 1.1 million subscribers. Members congregate here to talk shop about the market, share outlandish memes, and pepper each other with light-hearted teasing and support in equal measure. The community occupies an irreplicable space in the Reddit landscape with its instantly recognizable linguistic style, headline-catching antics, and vast reservoir of overzealous devotees. This paper will utilize Amy Jo Kim's community design principles along with social psychology theory as frameworks to understand how this chaotic, oftentimes offensive community has developed one of the largest and most loyal user bases on the platform. This paper will argue that humor plays a vital role in promoting in-group cohesion and in providing an unconventional third place for traders (and thinly veiled gamblers) to seek support from each other in the form of vulgar, yet good-humored taunting.

Before discussing what happens in WallStreetBets, it is necessary to take a quick detour to explain the subreddit's prefered method of trading: options trading. Options are contracts that give the owner of the contract the right to buy or sell (depending on the type of contract) an asset at an agreed upon price (strike price) within a specified time period. These are generally sold in batches of 100 contracts giving the owner the right to buy or sell one hundred units of the underlying asset. There's two types of contracts: put options and call options. Calls give you the right to buy the asset at the agreed upon price and puts give you the option to sell the asset at the



agreed upon price. More simply, if you think the price of the asset is going up, buy calls and if you think the price of the asset is going down, buy puts.

WallStreetBets users are notorious for buying cheap options that are nearing expiration and are far "out of the money" (i.e. an option that is worthless if exercised immediately). These are often incredibly cheap costing as little as a dollar since they are so unlikely to be profitable. These are in many ways like lottery tickets in that they most often result in losing the cost of the contract, but sometimes payout handsomely. What makes options so risky is that they are significantly more sensitive to changes in price of the underlying asset as each contract controls one hundred shares of the asset. This allows users to amplify their rewards, but also amplify their risk. A one dollar decrease in the price of a stock could mean a $100+ loss to someone holding a call option. The users of WallStreetBets play a game of big risk and big rewards, which oftentimes more closely mirrors the behavior of gamblers rather than investors.

High-risk trading is not the only defining trait of this group. It is important to note that this is a community that frequently engages in conventionally inappropriate behavior. This paper explores the depths of their "degeneracy," as some users have described it. Readers should be aware that this paper presents quotes and examples from this community, which feature uncensored homophobic and ableist language along with other generally vulgar expressions. This vocabulary is an important part of the culture and certain unsavory terms will appear frequently. Those who are uncomfortable with such language should proceed at their own discretion.



**Related Work**

Currently, it appears that there is no scholarly literature that directly focuses on WallStreetBets. As such, it is helpful to examine communities that exhibit a similar irreverence and disregard for prevailing standards of good community design. Pater et al. (2014) studied the community within the SomethingAwful forums using a participatory observation approach supplemented with semi-structured interviews. They focus on how SomethingAwful operates as a successful, vibrant community despite some surprising design choices such as 1) abuse of the newbie, 2) amplification of boundary pushers, 3) banning the boring, 4) public shaming, and 5) lack of internal consistency of moderation. The authors note that the success of these unconventional design decisions push forward the conversation of successful design features as being contextually dependent.

Bernstein et. al studied the role of anonymity and ephemerality in the design of 4chan (2011), a forum that WallStreetBets actively claims as an influence. They found that as a result of the anonymity and ephemerality users seemed to use slang and other linguistic cues to signal status and identity. Additionally, they argue that the ephemeral nature of the postings where content is rapidly deleted is shaping the community dynamics. They posit that the dynamics create a powerful selection method whereby content must be frequently reposted resulting in often increasingly creative mutations.

Given WallStreetBets users' trading style closely resembles gambling more than investing, it is worth exploring the boundary between investor and gambler. Statman (2002) investigated the shared kindrage of those who trade stocks and those who play the lottery. He



suggests that regular stock trading and playing the lottery are rooted in 1) belief that we're above average, 2) lofty financial aspirations, 3) motivation by fear and hope, and 4) an enjoyment of playing the game. Additionally, Mizrach and Weerts studied trading activity in an online chatroom where they claim that the traders in the chatroom "ignored many of the lessons from their finance classes" because "they trade very frequently" and "made no attempt to diversify," (2009) much like the WallStreetBets subscribers.

## Methods

Collectively, the authors spent over 180 hours of participant observation on WSB over the span of nine weeks, from January 27 to March 27 of 2020. They browsed the subreddit several times a day most days of the week. Their participation consisted of comments on new or popular threads at any given time. They posted a total of 131 comments combined. Only one of the authors attempted to post original content to the subreddit. They submitted two different memes on two different occasions, but the subreddit's AutoModerator[1] configuration promptly removed both of these submissions. Additionally, one of the authors participated in options trading following the advice of the subreddit. The authors also conducted semi-structured interviews with active users in the community. They catered the interview questions to the three types of users they interviewed: regular users, community moderators, and the community founder. However, each interview covered the same bases: the user's historical and routine engagement with the subreddit, their trading experience, and their opinions on specific characteristics of the community.

---

[1] https://www.reddit.com/wiki/automoderator



**Participants**

      The authors interviewed a total of 13 WSB users. The degree of involvement of these participants ranged from lurkers to moderators. All of the participants identified as male, which corresponds to the overwhelming majority of demographics in WSB. Due to time constraints and the scope of the class, the authors were unable to actively seek and recruit female or other minority participants. Plamen interviewed Allen[2], Nathan-T1[3], Ryan[2], Frank[2], and IS_JOKE_COMRADE[3]. Beatriz interviewed Max[2], haupt91[3], and Zack[2]. Christian interviewed ITradeBaconFutures[3], Mason[2], Henry[2], and William[2]. Both Christian and Beatriz interviewed Jaime Rogozinski. The remainder of this section is a description of each participant.

      **Allen.** Allen is 26 years old and he is from the United States. He works in the field of data science. Allen joined WSB in 2014 when he was in college and he checks the subreddit on a daily basis. Allen reads "new posts during market hours and read[s] the hot posts after markets close to get all the memes." Allen had very little trading experience prior to joining WSB and began trading options only after he joined WSB, shortly after "Robinhood added commission-free options trading." Allen does not take advice from WSB. He believes "most of the WSB advice is total trash or thousands of people regurgitating the same trade."

      **Nathan-T1**. Nathan-T1 is a 21 year old university student from the United States. He first encountered WSB in 2018. Nathan-T1 learned about WSB and joined the subreddit shortly after he began trading options. Initially, he checked the subreddit three or four times a day, but currently does not have the time to check the sub as frequently. He "usually just browse[s]

---

[2] Name is a pseudonym since this member asked to remain anonymous.
[3] This member asked to be identified by their Reddit username.



through and look[s] for some sort of fundamental/technical post that is a mix of humor and actual information." According to Nathan-T1, aside from a small percentage of users who understand trading, the trading advice given by the WSB community should not be taken seriously.

**Ryan.** Ryan works in the financial industry and lives in the United States. He has a college degree and is in his "late twenties, early thirties." Ryan joined WSB in 2017 and checks the subreddit two or three times a week, but seldom posts content. When he first joined the subreddit, Ryan noticed that WSB is characterized by its "pretty childish, crude humor, and surprising honesty in people's profit and losses." He initially subscribed to WSB because "people [were] posting their gains and losses and [were] just having some lighthearted humor at whatever [was] going on in the marketplace." Ryan traded options prior to joining WSB. He currently trades options, but does not take advice from the subreddit because he has "a strategy that works."

**Frank.** Frank is 31 years old and he lives in the United States. He is currently a full-time student enrolled in a master's degree program. Frank joined WSB between 2017 and 2018 and checks the subreddit two or three times a week. He does not post frequently because he "doesn't really have anything to contribute, as far as posting big, big bets or anything like that." Instead, Frank prefers to see what other people do. He subscribed to WSB because he liked the banter and "how people are sort of forced to put their money where their mouth is, so to speak". Frank had some exposure to trading options prior to joining WSB, and currently trades options. He does not take advice from WSB and thinks that users on the subreddit make risky bets.

**IS_JOKE_COMRADE.** IS_JOKE_COMRADE (or Comrade) requested that we mention his community flair which is "Bear Gang General." He is from the United States and



describes himself as a millennial who is currently working a "basic corporate job." The user joined the subreddit "this past Spring [2019], I believe. I knew of it for some time, that it was full of ridiculous memes and people that traded options." Comrade is heavily involved in the subreddit. Whenever he actively trades options he checks the subreddit every 20 to 30 minutes. He explained that "[WSB] is the 'front page' of what is going on that affects the market. Not CNBC, not CNN, it's here." Comrade had prior trading experience, but started trading options only after he joined WSB. He believes that WSB is "THE place to keep up to date with what affects stocks. Nowhere else. Plus, [he] get[s] good advice here. And it's really really fun."

**Max.** Max is a 29 year old professional engineer from the United States. He subscribed to WSB around late 2018 to early 2019. He checks the sub multiple times a day, and he began paying more attention to WSB after he got into options trading. By February of 2020, he had been watching, learning, and trading options for six months. Max began trading stocks after graduating from college thanks to his mother. Then he became interested in cryptocurrencies, which pushed him to self-study machine learning to automate trades. Max started his own trading meetup group, which he explains "really blew the doors off everything because [he] was exposed to and learning from legitimate market murderers."

**haupt91.** haupt91 (or Haupt) is 28 years old and is from the United States. He previously worked as a financial advisor at a major firm. Haupt joined WSB around late 2017. He started trading at the age of 18, and he first learned about the stock market through his father. Options trading has never been his primary method of investing, even after joining WSB. Haupt currently focuses on creating content for his YouTube channel, his podcast, and his subreddit–a spinoff of WSB. In the community, he holds the flair of "Bull Gang General" and he is known for his



memes. He posted his first meme on the subreddit around May 2018. By February of 2020, Haupt became a moderator of WSB. He often checks the sub multiple times a day because "the fluid nature of the stock market makes it so that people's opinions are interesting to hear about for why things are happening in between trading hours."

**Zack.** Zack is a 26 year old programmer from the United States. He first joined WSB at the start of 2019, and began trading shortly after they joined the subreddit. However, Zack's brother, who had been trading for a year at the time, was the one who got him into trading. He currently enjoys trading options. Zack considers himself a lurker and will rarely post on WSB. He believes that, like him, the majority of users will visit the subreddit without commenting or posting. "I go there to read about funny stuff, not really [to] talk to people about it. I don't know. I keep my research to myself usually."

**ITradeBaconFutures.** ITradeBaconFutures (or Bacon) is a technology professional in his late thirties from the United States. He believes that he first started participating in the subreddit around 2012 after watching the subreddit speculate on "weed stocks" at a time when marijuana was still completely illegal. Eventually Bacon was promoted to moderator and is currently a senior moderator for the group. He describes the subreddit as "one of the most fun places that you can't ever tell your family or professional circle that you're involved with." Bacon boils down the philosophy of WallStreetBets to "Let's go make some fucking money. Let's not sit here and talk about what my 401K's going to give me when I'm 65." He claims to have developed many meaningful relationships with other users in the community and even claiming one such relationship helped him "evolve as a human." Additionally, he commented "that's the kind of relationship and bond you can form with this group." Currently, he spends his time as a



moderator less wrapped up in the day-to-day operation of the community and is more so focused on thinking up "new ways to expand the platform."

**Mason.** Mason is a 24 year old grocery store employee from the United States. He first encountered WallStreetBets in 2018 when he was introduced to the subreddit by his brother who described it as a "place where people go and lose all of their money." Prior to joining WallStreetBets, he had never traded options and found the group to contain a wealth of "great information" to get him started trading. Currently, he feels the useful information is a bit harder to come by and that the subreddit is dominated by "shit posting." Despite this claim, he still checks the subreddit daily using it mostly to find information about earnings calls and for his own entertainment.

**Henry.** Henry is a French Canadian in his 30s working in the public service sector. He first encountered WallStreetBets around 2018 and was taken in by the "pretty awesome memes [*sic*]" and the fact that WallStreetBets users "had a very different take on the market" involving a "big risk big gain" orientation. He claims to be on WallStreetBets "basically non stop" during trading hours and sees the website as an excellent aggregator of financial news. Henry even goes so far as to say that "I strongly feel i'm getting a more accurate picture of the reality on wsb than on mainstream media." Despite his grievances with the recent influx of "useless comments," the "unique trading culture" he believes is inspired by films like *The Wolf of Wall Street* and *The Big Short* keeps him coming back.

**William.** William is a 22 year old telecommunications engineer from the United States. He first encountered the subreddit around 2018 and was captivated by what he described as a "clusterfuck and cesspool of half idiot half geniuses all trying to get rich." As of recently, he



mainly uses the subreddit as a source of entertainment and believes that the quality of trading advice has declined in recent months describing it as "horrible" advice given by people who have "absolutely no idea what they're talking about." Although William doesn't care much for the advice shared on WallStreetBets, he comes back for the "sense of community" and claims WallStreetBets users cultivate this sense of community by virtue of them "all trying to reach the same goal by playing the same game." His admiration of the community runs so deep that he described his first encounter with WallStreetBets as feeling like he "was home."

**Jaime Rogozinski.** Jaime is a 38 year old entrepreneur from Mexico. He is the original founder of WallStreetBets, which he created in 2012 imagining a " place where people could come up with profitable, consistent, responsible day trading strategies." Jaime mostly uses the subreddit for entertainment and feels responsible for guiding the community. He acknowledges that this guidance often "comes through the moderators," rather than his own voice. When reflecting on the status of the sub today, Jaime explained that "[he] can't take credit for having made what you see when you go on the sub today. The community made it; the moderators made it." Jaime recently wrote a book focusing on the role of WallStreetBets in the current financial ecosystem titled *WallStreetBets: How Boomers Made the World's Biggest Casino for Millennials*.

## WallStreetBets As A Third Place

It is helpful to view WallStreetBets as a "third place," as defined by Oldenberg in his 1999 work *The Great Good Place*. Oldenburg describes third places as being separate from the first and second places of home and work, respectively (1999). These third places also embody



several additional characteristics: (1) They are neutral ground, (2) They are leveling places, (3) conversation is the main activity, (4) They are accessible and accommodating, (5) The regulars set the tone, (6) They maintain a low profile, (7) The mood is playful, (8) They are a home away from home. WallStreetBets appears to fit quite neatly into this foundation.

**A Neutral, Leveling Place**

WSB provides a neutral, level playing field where users are free to join and post regardless of whether they hold graduate degrees in finance or are an 18 year old that just downloaded the Robinhood brokerage app. Per the site's founder, Jaime, the forum is populated with a mix of users who are "experienced and have different levels of risk tolerance" and users who are "completely new" and have "no concept of risk whatsoever." Jaime further claims that, despite the vast experiential divide amongst many of the users, they find common ground in that they "have the same goal of making money." Several other participants echoed this sentiment that there are folks of all experience levels with one even describing the user base as equal parts "idiots" and "geniuses."

Oldenburg also notes that in third places "those not high on the totems of accomplishment or popularity are enjoined, accepted, embraced, and enjoyed despite their 'failings' in their career or the marketplace." One could almost imagine Oldenburg saying this upon viewing WallStreetBets, a place where people of all backgrounds come together to champion those who literally "fail" in the marketplace. This celebration of losses is so pervasive that there is even a rule banning this behavior on the subreddit. As Oldenburg argues, "Even



poverty can lose much of its sting when communities can offer the settings and occasions where the disadvantaged can be accepted as equals."

Additionally, there is a potable sense of camaraderie in WallStreetBets where users celebrate the entirety of the trading experience. Statman (2002) notes that "camaraderie is the rule in casinos even at many games in which players compete" because players are "united in action." Going further, Statman draws similarities to day traders saying that "camaraderie is the rule in trading rooms" and that "although traders compete with one another, they also cheer one another." Perhaps these parallels offer insight into the leveling practices of WallStreetBets.

**Conversation is the Main Activity**

It may seem from the outside that the main activity of WallStreetBets is options trading, but several participants have made it clear that the real value of the group is derived from the playful banter. This banter usually manifests in the form of humorous memes and shitposts. Bacon even claimed, "they spend more time analyzing comments and coming up with really good shitposts than they do actually looking at good stock points." Another prolific user, haupt91, advises that if you plan to post, you should not be "repetitive" and "always try to be unique." These claims seemed to be congruent with Oldenburg's assertions that speech in third places is "more often attended by laughter and the exercise of wit."

Perhaps one of the reasons why WSB is so successful is because new conversations are constantly developing. This is mainly due to the ephemeral nature of the stock market. The memes and jokes created in response to a particular stock or overall market trends may be drastically different throughout the day. In addition, because of the inherent unpredictability of



the market and the innumerable factors that affect it, you may never know what will be popular on the subreddit within the next day or even within the next hour. Bernstein et al. made a relevant observation in their study of 4chan, which is a large community characterized by ephemeral content. The authors suggest that part of the reason why 4chan is so popular is because the ephemerality of its content encourages users to participate.

Additionally, Oldenburg suggests that conversations in third places are "more spirited than elsewhere, less inhibited and more eagerly pursued." This could perhaps explain the subreddit's constant "trolling" as some users put it and the lack of inhibition in the use of words like "autist" and "retard" that they would not likely use so freely in their daily life. Henry, when asked why WallStreetBets maintains this controversial vocabulary, suggested imagining "a private bar discussion where you wouldn't hold anything back." It is also expected that, despite the linguistic inhibition, the mood is playful and non-aggressive. One user, Frank, described participating in WSB as "similar to play[ing] on a sports team" or like being part of a fraternity because "you get made fun of a lot" and "get called up all these different names," but "it should not be taken seriously."

**Accessibility and Accommodation**

Oldenburg also highlights the importance of accessibility and accommodation in constructing a functioning third place. Oftentimes, WallStreetBets can be far from accommodating to new users with most veterans complaining about them and some interviewees even going so far as to describe them as "cancer." The interviewee that took the leap to call those new to the subreddit "cancer" suggested that they will be embraced once they are "familiar with



the culture of the subreddit." WallStreetBets appears to be accommodating to those that respect

the culture and have the expected cultural domain knowledge. Once these are in place, users are

largely accepted in a place open 24/7 where some of its nearly 1.1 million subscribers are surely

kicking around.

**A Low Profile And A Playful Mood**

One of Oldenburg's criteria to be a third place that WallStreetBets seemingly subverts is

that a third place maintains a low profile. Oldenburg claims that "third places are unimpressive

looking for the most part," yet WallStreetBets is extravagantly designed with a dynamic banner,

a color scheme incorporating the entire rainbow, and eye-poppingly crude language. But, the

truth is WallStreetBets maintains a low profile through what Oldenburg describes as "protective

coloration." Oldenburg views the plainness of third places as aiding in not attracting "a high

volume of strangers or transient customers." The outlandish and oftentimes offensive culture of

WallStreetBets creates the same effect.

Something plain and with proper language may be attractive to someone who wants some

responsible investing advice, but a place where users call each other "retards" and celebrate the

loss of thousands of dollars will likely be off-putting to the majority of retail investors. Jaime

argues that this vocabulary makes WSB "feel like a closed community even if it's a million

people strong." Oldenburg gives a similar example of this technique in the physical world saying

"A place that looks a bit seedy will usually repel the transient middle-class customer away." He

also goes on to say that "a definite seediness still goes a long way toward repelling the female



customer." Perhaps, this could begin to explain why all thirteen participants we spoke to were male.

It seems that the users are very well aware of this "protective coloration" that the subreddit employs. One user (Henry) said that the harshness embodied in the language and user interactions helps to "intimidate and purge" people not in the right mindset and Jaime claims "a lot of the outsiders can come in and feel unwelcome." However, several users claim that once users get past the language, there's a lot of value to be found here. One user (IS_JOKE_COMRADE) claimed that it took him a while "to realize that the level of thought here is (occasionally) quite sophisticated" and another claimed that he quickly realized underlying the hijinks "there was some intelligence there." (Bacon). This underlying intelligence fits well with Oldenburg's description of the utility of a low profile for third places. Oldenburg states that the low profile of "the third place is all the more likely not to impress the uninitiated." Bacon argues that users who are "able to parse through the intent versus the actual verbiage" of the content on WallStreetBets are more likely to get something out of it and stick around.

Another function of the plainness of third places Oldenburg highlights is that it serves to "discourage pretension" among those who gather there. Many of the users expressed in some form or another that mainstream financial institutions and media often take on an overly self-serious tone. One user (haupt91) described the serious tone of the media outlets as "so serious" that "it's like someone died," and Bacon described the people on CNBC as "screaming with their hair on fire." Most interviewees contrasted the tone of these outlets and more serious financial subreddits (r/finance, r/investing, r/stocks) with that of WallStreetBets, which they described as the "antithesis of a serious subreddit" (Bacon) with a whole culture built around the



idea that "nothing is to be taken too seriously" (William). Oftentimes, users will mock people that offer conservative or overly serious investment advice telling them to "go back to r/investing" or calling them a "boomer." This seems to concur with Oldenburg's observations that those that come "overdressed" to third places will likely be met with "a good bit of ribbing." Furthermore, it exalts another popular practice from SomethingAwful, where "the boundaries for what is considered 'shitposting' and what is considered valuable contributions" are fuzzy (Pater et al., 2014).

Additionally, WallStreetBets appears to have a distaste for serious retail and institutional investors who believe that their investment success is the result of their own merit as an investor. One user describes retail investing and trading as "a giant game that's rigged against all of us," whilst another user praised the 1973 book *A Random Walk Down Wall Street,* which posits that stock asset pricing exhibits behavior similar to that of a random walk and thus cannot be predicted. Jaime explains that on other forums users are "bragging" saying things such as "I'm right, this is what is going to happen with Apple." Meanwhile, WallStreetBets users actively undermine their own success by referring to each other as "retards" and attributing their success to their facetiously self-diagnosed autism. For example, one user posted a gain from a trade that had his account worth over ten million dollars and titled the post "NEW HIGHSCORE!!!" treating a return on investment that would make most hedge fund managers green with envy as if it were simply points in a video game. The responses to the post were often equally as flippant with one user commenting "we've entered the era of the eight-figure retard"[4].

---

[4] https://www.reddit.com/r/wallstreetbets/comments/fgg4fg/new_high_score/



It appears common that regardless of whether a user loses or gains vast sums of money they will be labeled as an "autist" or "retard" by the community if they have not already done so themselves. There appears to be a duality to these terms. If a user makes a big bet that ends up making money they are praised as "autistic" in the same sense as the titular Rain Man from the 1988 Dustin Hoffman film. Rain man was able to use his recall abilities as an autistic savant to count cards and make a fortune gambling, but despite his calculation abilities he was still disabled and unable to live without assistance. If a user makes a big bet that ends up losing money, they are still characterized as being "autistic" or "retarded." The only difference now is that the term holds no backhanded praise and simply labels them in a very derogatory way as an idiot. In a way this kind of diminutive language acts as a leveler by saying in offensive terms that essentially "we're all idiots here," but some idiots are luckier high-risk gamblers than others. This also seems to have the added utility of keeping the mood playful by not taking any user's performance on the market all too seriously.

Research shows that an effective way of increasing adherence to a norm is to publicly contrast inappropriate behavior with a descriptive example of a norm along with corresponding appropriate behavior (Kiesler et al., 2012). As was mentioned earlier, the low profile of a third place is meant to "discourage pretension" and that those "who should arrive overdressed" should expect to be met with "a good bit of ribbing." This holds true for people who come to WallStreetBets with serious advice about a certain security. As one user puts it "if you have an idea, you claim to have a position, you will get huge shit whenever you don't show proof of your position" (Frank). It is expected that if a user has a certain belief about movement in the market they should "put their money where their mouth is" (Frank) by staking out a position based on



their own advice. Users who do not do this are often ridiculed as being inauthentic or pretentious for offering advice they wouldn't take themselves. The most common way that this manifests is in users responding "positions or ban" to other users' claims, which acts as an implicit request to the moderators to make this person prove that they have something riding on their claim or ban them. Most claims on WallStreetBets are not taken seriously, but if one wishes to have theirs taken seriously, they must be willing to take on large amounts of personal risk to prove it. There is simply no room for pretenders or as the subreddit rules put it "paper trading teenagers." The reliability of assessment signals, as described by Judith Donath (2002), is priceless in WSB. Anyone can say they support a specific trading position, but if you can demonstrate that you actually invested in that move, then users might begin to take your suggestions seriously, and even act upon it as well.

**The Regulars**

Another important attribute of third places that Oldenburg notes is that the regulars set the tone. Oldenburg states that "it is the regulars whose mood and manner provide the infectious and contagious style of interaction" and this is something that holds true on the subreddit. Several interviewees mentioned users like haupt91 who claims to spend hours making memes that are known throughout the community for their influence on the humor. Haupt's memes even caught the attention of a Forbes Magazine contributor who claimed haupt91 was responsible for some of the subreddit's "most creative content" (Alfonso, 2019)[5]. Bacon claimed that users like haupt91 displayed commitment to creating memes and shitposts "that really showed their creativity and their ability to think about what the topic is" and that these contributions "resonate with the community." Perhaps this resonance across the community is similar to the resonance of the witty banter of regulars amongst patrons at the third places Oldenburg observed.

---

[5] Note that haupt91's age and location are fictitious in this article. Haupt explained during a WSB-related livestream that this was a precautionary measure for when he used to work in the financial industry.

There are other users who have left their own mark on the community in their time on the subreddit, such as 1RONYMAN, who famously misused a trading strategy called box spreads. He claimed on the subreddit that his technique "can't go tits up" (i.e. can't fail), but subsequently lost more than $50,000, which in turn led to Robinhood banning the use of box spreads on their platform[6]. Nowadays, users will often joke about outrageous positions and sarcastically claim that their strategy "can't go tits up," honoring the history of the community that generally only regular members will comprehend.

Additionally, there are users like ControlTheNarrative who famously found a glitch in Robinhood that allowed him to borrow unlimited amounts of money from Robinhood on margin. He proceeded to bet a large amount of this borrowed money against Apple before earnings, after which Apple's stock went up, and left Robinhood on the hook for the $48,000 he lost. Most notably, he live streamed this process and when he realizes that he lost all this money his only response is the guttural sound "GUH." Now, users will often comment the phrase "GUH" in response to any post or comment about a user losing money.

Regular users like haupt91, 1RONYMAN, and ControlTheNarrative have helped build the vocabulary and stock the joke repository of the community. As Oldenburg suggests, the comings and goings of regular users like these are noticed by the entire community. Most recently the accounts of haupt91 and ControlTheNarrative were removed from Reddit and users instantly noticed it with posts cropping up about these losses receiving thousands of upvotes and innumerable mournful GUH's. The highly publicized actions of these users have shaped the tone of interaction on the subreddit whether it be in setting an example for the quality of content and shenanigans or simply leaving a mark on the WallStreetBets lexicon. Either way regular users like these continue to help evolve the culture of WallStreetBets.

---

[6] See "Why aren't box spreads allowed on Robinhood?" at
https://robinhood.com/us/en/support/articles/360001331403/options-investing-strategies/

**A Home Away From Home**

Finally, Oldenburg notes that third places have the quality of being "homes away from home" to those that frequent them. Many participants expressed great fondness for the subreddit with one even explicitly describing his discovery of WallStreetBets saying "I felt like I was home" (William). Oldenburg describes third places as ideally suited for social regeneration where users can proverbially "let one's hair one." Most all of interviewees expressed that they were actively involved in day trading, which was regularly described as "repetitive" and "a boring job" (Jaime) that often just consists of "watching a screen all day long" (Henry). Many users described WallStreetBets as offering them a little bit of entertainment and engagement throughout the day with one user describing the subreddit as bringing "some fun along the way" in an otherwise boring day of active trading. Jaime echoes this statement describing WallStreetBets as an "outlet" for professionals "in front of their computer all day long doing these repetitive things over and over" where "it's entertaining, it's fun, it's refreshing."

The sense of a "home away from home" provided by third places is often facilitated by the "freedom to be" that is manifested "in conversation, joking, teasing, horseplay, and other expressive behaviors." A moderator of the subreddit (Bacon) stated that WallStreetBets users



should "feel free to make a joke that [they] think is risky, that [they] can't make in [their] personal life." However, he also notes that they do moderate posts they believe are genuinely prejudiced as he states "believe it or not, none of us are anti-Semitic and none of us are homophobic." Another interviewee (Henry) echoed this sentiment stating that the "tolerance is pretty high here" in regards to the jokes and language used "as long as it's stock related and not getting on or targeting a specific individual." In keeping the spirit of the subreddit playful, it seems that users are given a lot of leeway to feel uninhibited in their language and jokes so long as it is understood by the community that they are not being serious. As Bacon states, "I think really the thing we imply the most is just humor will excuse a whole lot of transgressions." Just as in SomethingAwful, WSB gives "[its] users a public platform to push the boundaries, [where] users are celebrated and rewarded for their efforts" (Pater et al., 2014).

Finally, Oldenburg notes that one of the most important things in making a community feel like home is warmth. Oldenburg describes warmth as emerging out of "friendliness, support, and mutual concern." Jaime claims that talking about a topic as intimate as money helps foster a "sense of unity" among the community. He states that people on WallStreetBets can openly say something like "I just lost ten grand," which "you can't tell your wife or you can't tell your friends or your mom" and that there is a certain "therapeutic aspect" in strangers on the internet telling you "you'll get through it." Jaime goes further to state that some users will say things about the community like  "you took me out of my depression" or "you truly are like my second home."

Bacon claims that in his experience behind the veneer of obscenity, generally "everyone's really generous and kind" and "willing to share their expertise to help you." He argues that often



times the crass language is used in a fashion of mutual concern where someone is "trying to help you by calling you a dumb ass or an autist" in order to help you "get past your own stupid biases" towards the market. Overall, it appears that many users do consider WallStreetBets a second home and feel a deep sense of belonging.

In review, it seems that WallStreetBets acts as a third place for mostly young traders to freely joke and associate in ways that they cannot in their day to day lives. We can see all of the eight criteria of third places set out by Oldenburg seem to be represented in the WallStreetBets community. (1) It is a neutral ground where users are free to come and go as they please with no obligation to the community. (2) It is a leveler where traders of all backgrounds are part of the chaotic conversation. (3) Conversation is the main activity with users competing for the best shitposts and memes. (4) It is accessible and accommodating to all who are willing to respect the culture. (5) The regulars set the tone shaping the humor and language of the community. (6) It maintains a paradoxically "low profile" with its antics detering those unwilling to look deeper. (7) The mood is playful and nothing is considered too serious to joke about. (8) It is a home away from home with many users finding a sense of community and belonging. To its users it is an undeniably important place, and we hope to further understand how the design of the community facilitates this through the lens of Amy Jo Kim's community design principles.

**Evaluating WSB Design Principles**

We selected the most relevant principles from Amy Jo Kim's *9 Timeless Design Principles for Community-Building* (2000) to evaluate the design of WallStreetBets. In the following sections we will answer the questions outlined below with respect to this community.



Does WallStreetBets:

1.  Have a clear purpose?

2.  Have distinct, extensible gathering places?

3.  Have profiles that evolve over time?

4.  Have a clear-yet-flexible code of conduct?

5.  Organize and promote cyclic events?

6.  Provide a range of roles with increasing involvement?

7.  Integrate what's online with the real world?

8.  Promote effective leadership?

**1. Purpose**

WallStreetBets was founded by Jaime Rogozinski in 2012 with the purpose of sharing "high-risk investing or trading ideas" (2020). When we spoke with Jaime, he explained that he knows of no other place on the Internet where people share their opinions towards the stock market as openly as they do on WSB. He criticizes other forums for being places where users brag and inflate their egos, whereas users on WSB joke and have fun because it has become a "humorous outlet." However, Jaime reflected that the subreddit has steered away from the intended purpose he once envisioned.

> "I wanted it to be a day trading sub. I wanted it to be a place where people could come up with profitable, consistent, responsible day trading strategies. And it went away from that, but what we got instead is actually really cool, too, in its own way. And it's really special. And I'm really happy with what it turned out to be, but a lot of the things that I



did want did end up continuing, which was a fun place, which is honest, where people are nice to each other."

Although it may seem counterintuitive to view WSB as a venue for support or to find nice people, all of the participants do perceive these positive qualities. For instance, Max explained that he enjoys "the hilarious posts about the markets that are super relatable as [he] experience[s] the market in [his] own way." He also added that WSB has given him "the chance to talk to actual professionals, pose questions, and get feedback that [he] would otherwise have no way to get." Another participant (Allen) described the appeal of WSB for him is that "the community is raw and people make a lot and lose a lot, and aren't ashamed." Henry explained that WSB provides users with "a sense of community" and "a shared goal" because they are all "playing the same game" and "[they] like to know there [are] other people who are the same as [them]." A common sentiment among the participants was their appreciation of the humor in WSB. Zack asserted that "when [they are] reading through some of the comments, [they are] laughing [their] ass off." Figure 1 is another example of how much WSB users appreciate the lightheartedness and the humor in their community.

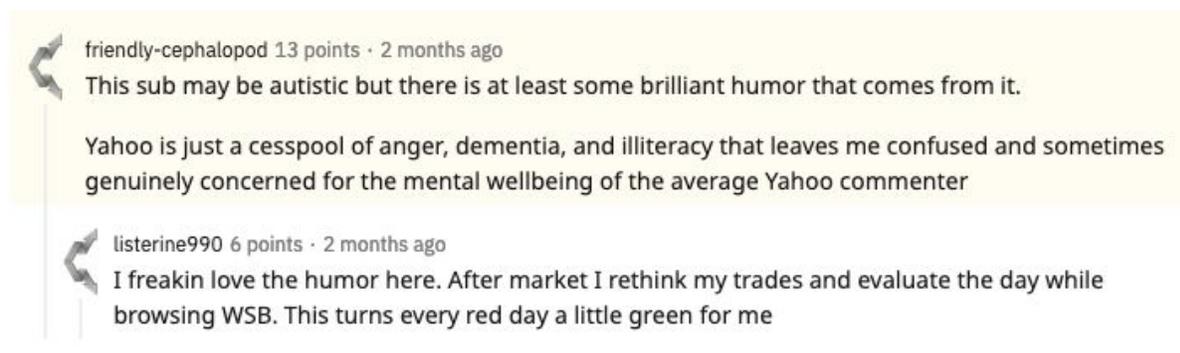

Figure 1: Two WallStreetBets subscribers discuss the advantages of their community over Yahoo Finance.



**Audience.** The quintessential member of this subreddit is a millennial, anglo male. When we asked our interviewees to describe the audience of WSB, six of our interviewees, including the moderators and the founder, speculated that most of them are college students. Most participants also believe that the members of WSB might also be engineers, working professionals, and oftentimes even working in the finance field. When describing this audience in general terms, Max explained that WSB is composed of "retail and professional traders, and fans of comedy." He adds that it is also a "fairly intelligent crowd, but still littered with complete dumbasses." IS_JOKE_COMRADE's observations concur with Max's, for he believes that WSB users are "college educated for the most part, but not always–probably on the higher end, though." It is also helpful to note that all of our interviewees fit this description.

During our interview, Jaime explained that WSB is a place where people with different backgrounds and risk tolerances, as well as different approaches to the market, congregate. In particular, Jaime expressed that "[this] junction is just chaotic." Jaime wonders "how can you have an organized discussion with two people that don't want to understand the same thing, but have the same goal of making money?" When addressing the general behavior of this audience, Haupt laughs and explains that "risk-tolerant people would be the biggest uniting personality trait" among subscribers of WSB. None of the participants believe that there are any particularly discernible sub-groups within this audience. According to Haupt, the only categories of note "are people that just have an opinion on which way the market's going to go." Nonetheless, he "[doesn't] think there's really a clique of people besides maybe the mods, who are notoriously gay." Part of the culture of WallStreetBets is that users cannot talk about the moderator team without making a homophobic joke. Next we will reiterate some of the discussion on the



language in WSB with respect to the purpose it serves for the audience. We will specifically address the homophobic jokes aimed at the moderators under the section on the code of conduct.

In his book on WallStreetBets, Jaime highlights that members of the subreddit "created their own lingo and often use crude, offensive language with complete disregard for people's political stances, religion, nationality, gender, sexual identity" (2020). He also adds that "WSB members are unbiased, equal opportunity insulters." Bacon explains the origin of this language as coming from "people that don't want to take stuff seriously and want to have that safe space to just be an asshole sometimes." In particular, Bacon argues that this works because WSB members can "be an asshole with some friends that are going to give you the same amount of shit that you give. And we're all going to laugh about it at the end of the day."

**Mission.** The subreddit's Frequently Asked Questions (FAQ) wiki[7] introduces WSB as follows.

"/r/wallstreetbets is a community for making money and being amused while doing it. Or, alternatively, a place to come and upvote memes when your portfolio is down."

From the outside it does not seem like WSB provides much value to its members' lives. In fact, it often seems like the opposite: a foul-mouthed flame fest where you get shamed for stupid questions and everyone loses significant sums of money. Jaime explained to us that for professionals day traders, WSB serves as "an outlet where it's entertaining, it's fun, it's refreshing," which they can take a look at while they sit in front of their computer performing repetitive and menial tasks. Bacon also explained that in order to appreciate "the content [in WSB] for the context and the larger undercurrents of what they're trying to communicate" that

---

[7] https://www.reddit.com/r/wallstreetbets/wiki/faq



readers must "have a certain amount of intellectual capacity" and "think critically" in order to

not be turned off by "the heinous stuff they say." He also acknowledges, however, that the

content in the subreddit is not palatable to everyone, and that some people have an "emotional

reaction and then they walk away." It is worth noting that to a WSB member, what appears to be

indiscriminate flaming actually has a "subtext" where users are "actually trying to help you by

calling you a dumbass or an autist, because you couldn't get past your own supid biases [on the

market]" (Bacon).

It should not, however, come as a surprise to learn that WSB does serve a purpose for its

members. Users are uniquely passionate about their home away from home and they are not

afraid to show it. For instance, as seen in figure 2, users recently panicked around mid-March

when the subreddit reached one million subscribers because the moderators made the subreddit

private as a prank in response to the milestone.

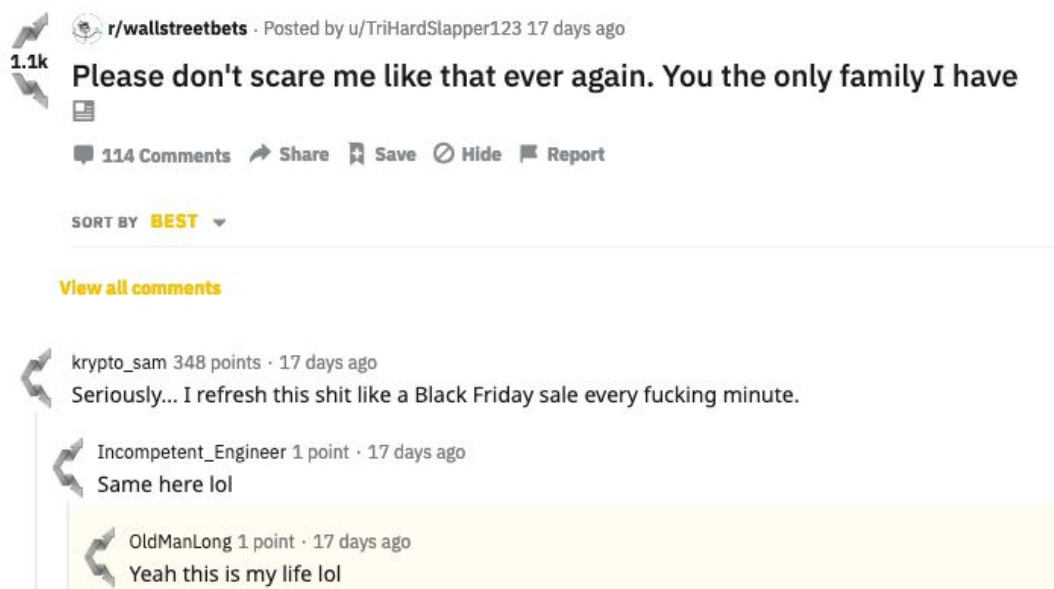



Figure 2: WSB users lament over the incident when the moderators briefly took the subreddit

private the day in which they reached a million subscribers.

Independently, Jaime, Bacon, and Haupt suggested that there is more to WSB than meets the eye. Eventually the endless stream of memes and jokes as well as the users' cynicism led the community towards an almost activist attitude. When Bacon considered a particularly crude and pervasive joke that was once popular in the community, he explained "Why? I don't know, it's just rainbow dicks was a thing we did. It became the thing for 'now we were trying to seriously troll the actual establishment of finance.'" And it is in fact this trolling, which will deter many unsuspecting outsiders, which allows them to respond critically to their experiences. In particular, haupt91 argues the following.

"There's something to be said for what our community allows people to do. What it

represents. The finger in the eye of this fake meritocracy that we live in and this

economy. There is some depth here–is my point. It's not just memes. I think that we do

put our finger on the pulse of the irony of our current economic situation in this country."

And this very chaos, yet uncanny sensitivity, of WallStreetBets allowed Jaime to write his book, which he explains is "not about the subreddit." In fact, Jaime "used examples from the subreddit to illustrate [his] point," which is to highlight "systemic risks" of the stock market "that people may not realize exist."

**Visual design.** The visual and stylistic choices on WallStreetBets effectively reflect on the community's lightheartedness and irreverence towards the stock market. The most notable visual trait of WallStreetBets is its conspicuous decor and attention to detail. Although Reddit as a platform limits the amount of customization that can go into any page, WSB takes advantage of



these features as much as possible. For instance, in the old Reddit view[8] posts are color-coded based on their flair with large icons and bright, flashy colors (see figure 3 for an example). In the "new" Reddit, the up and down arrows for posts and comments are customized to resemble simplified or cartoon-ish versions of stock market trends, as shown by figure 4.

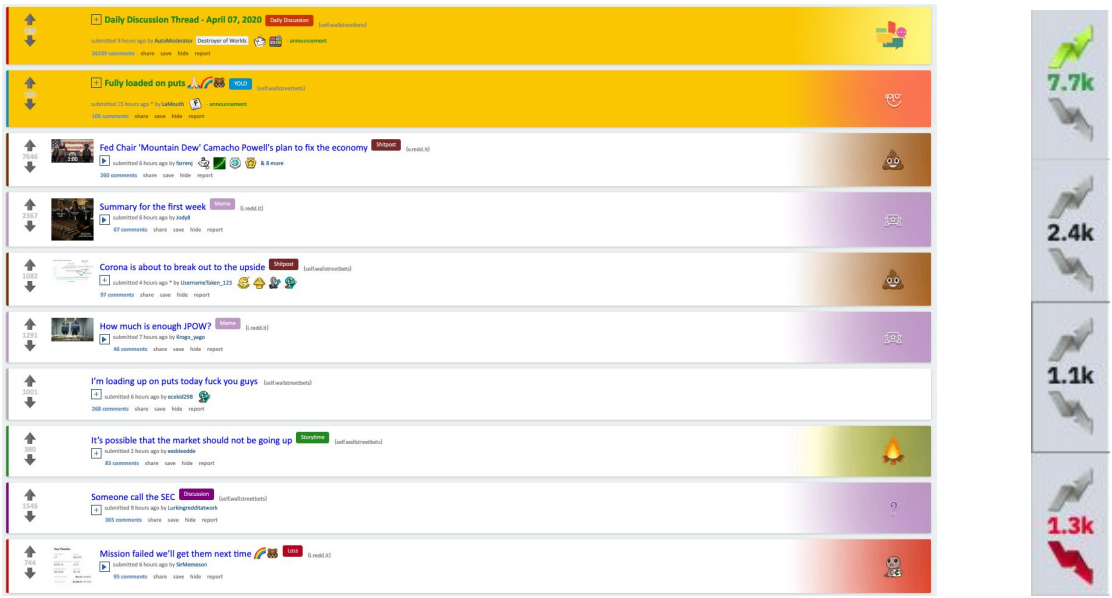

Figure 3 (left): Front page of WallStreetBets on old Reddit view.

Figure 4 (right): Example of the up and down arrows in WSB.

When you navigate to WSB from the browser, the most relevant links (e.g. community FAQ and background information on finance terms) are visible at the top, along with a dropdown menu that allows you to filter the visible posts by "flair" or category. WSB users expect you to learn, lurk, and have a basic understanding of finance (particularly options trading) before you begin to post, thus it makes sense that these sorts of resources are appropriately highlighted within the community. Bernstein et al. made similar observations about the 4chan culture, where





the "lack of fluency is dismissed with the phrase 'LURK MOAR,' asking the poster to spend more time learning about the culture of the board" (2011). Furthermore, Nonnecke and Preece discuss that the rapid growth of a community could "create chaos and lurking in large [discussion lists] may be a practical means of reducing the number of posts and maintaining order" (Nonnecke & Preece, 2000), which aligns with the beliefs of the participants we interviewed.

Unlike many other finance-related subreddits, WSB has its own logo, which reinforces the community's personal brand and also serves as their mascot. Until March of 2020 their logo was some variation of a cartoon baby wearing sunglasses and a suit, which the community affectionately referred to as "fuck boy." This character was often depicted with its arms raised with a motion that simulated throwing money in the air (see figure 5). Due to copyright concerns, the WSB administrators updated their mascot to a new baby character, which the community quickly denounced. For example, in one of the many threads that discussed the logo change one user commented "Who thought this would be an improvement? The old logo was a staple of this sub and changing it at all (especially to this shit) just waters down the WSB brand." One user even took the time to write a script and post a tutorial to the subreddit on how to replace the new logo with the old one, which works on both old and new Reddit site customizations.

The backlash over the new logo was so significant, that the subreddit moderators changed it in favor of a poorly drawn stick figure (see figure 5), for which the community was ironically thankful. One prominent WSB user created a "Daily New Logo Appreciation Thread" for the stick figure replacement in which they claimed that "there are no complaints about it and it's far more popular than that atrocity we were exposed to over the last couple of weeks."



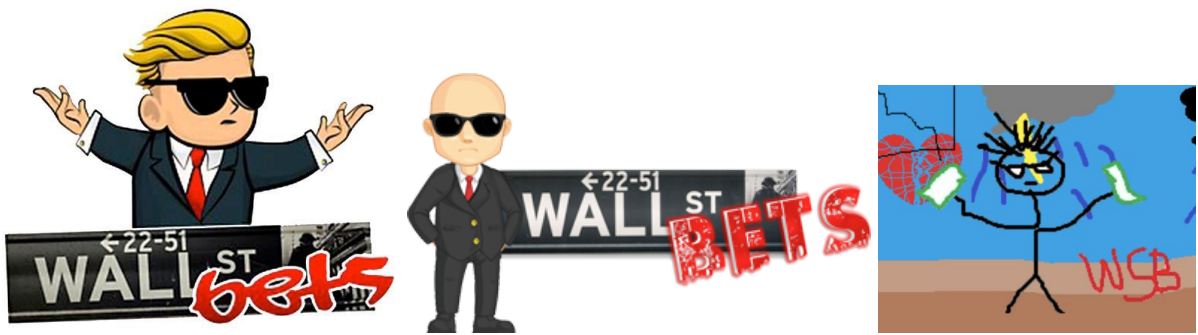

Figure 5: Old WSB logo (left), the newest logo (center), stick figure (right) .

WallStreetBets also features a dynamic banner image, which displays some basic animation whenever you hover your cursor over the image. The banners will also change based on the season or current events. For example, during the Chinese New Year, the baby character was seen wearing a traditional Chinese robe instead of its typical western suit (see figure 6).

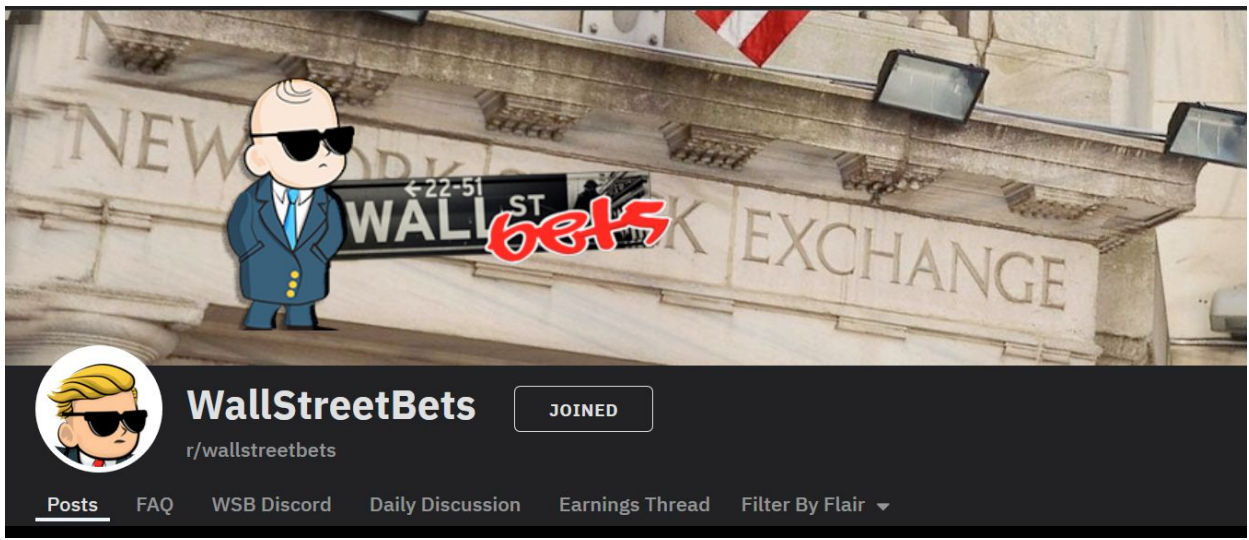

Figure 6: Screenshot of the WallStreetBets banner and logo from the new Reddit view, courtesy of u/iamchairs.



On the other hand, other finance subreddits like r/investing–which is a place where WSB users will send you to if you ask a stupid question–are completely lackluster in comparison. These rarely present any custom styles, and even keep the default "profile" image of a planet (see figure 7). The few images they do use to adorn their pages are often static, unchanging, and reflect a particular point of view towards the market. This presentation reflects the more serious tone of the finance industry, which participants such as Jaime, Bacon, and haupt91 alluded to during their interviews.

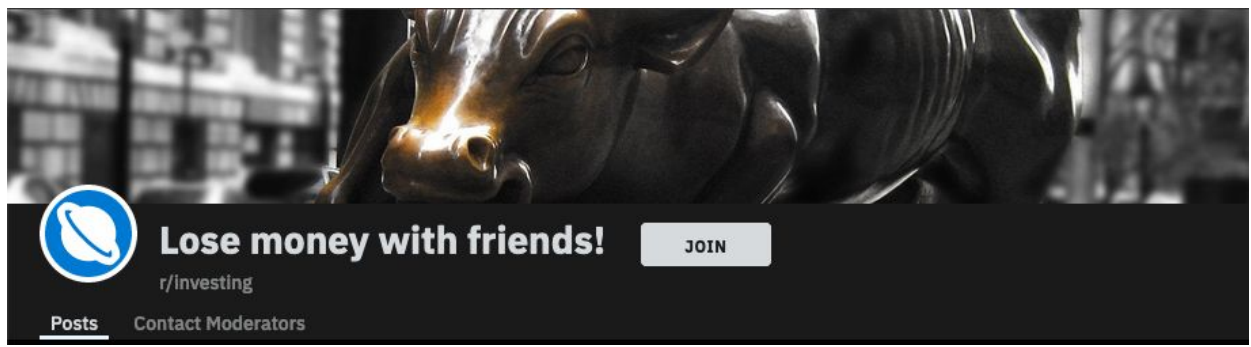

Figure 7: Screenshot of the r/investing banner from the new Reddit view.

Finally, user flair, which is often used to help fellow members easily discern sub-groups within any subreddit, is also "abused" in WSB. Although some users will receive appropriate flairs that properly indicate their particular status (more on that under the Roles section), plenty of the users display humorous and sometimes raunchy flair. Sometimes they even add unnecessary symbols and emojis. In particular, the most ridiculous flairs are typically awarded to the moderators, as can be seen in figure 8.

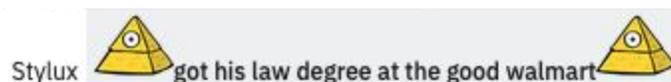

Figure 8: Username and flair of one of the WSB moderators, u/Stylux.



## 2. Gathering places

**Places.** Though many of our interviewees say that they do not congregate anywhere in particular and are content to scroll through the front page of the subreddit, several interviewees mentioned the role of daily recurring threads such as the moves thread and the daily discussion thread. These will be covered more explicitly when discussing cyclical events, but they seem to play the role of allowing a sidebar discussion for topics that are not important enough to merit their own post. Users can come here to find live commentary on the market, talk about their positions, or simply crack a few jokes with other users. In a way, they serve as the proverbial "junk drawers" of the community in that if you're not sure where something belongs, you can always throw it in here with no consequences. This principle also appears to hold true for the subreddit's relatively new Discord channel where users are free to discuss with little focus and can receive immediate feedback.

**Map.** As has been discussed previously, WallStreetBets is a place where satirical content and well-researched investing advice are presented side by side, often appearing indistinguishable to the untrained eye. To make it easier to understand what is what, the subreddit tags all posts into one of several predefined categories. On the side bar, users are able to filter through posts depending on their tags as can be seen below in Figure 9.



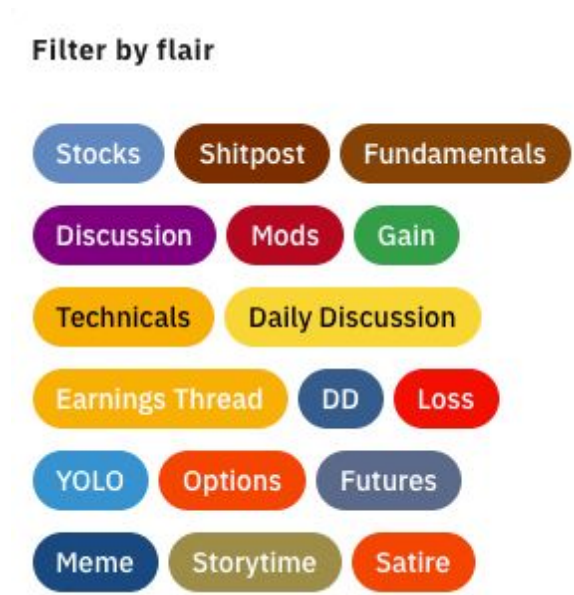

Figure 9: Post Flair Filter

        This tagging system allows users to be able to understand what kind of content they are

viewing. It also gives users the opportunity to self-curate content based upon their individual

needs. Some users may simply be interested in the entertainment value of the subreddit and are

free to look at content that is categorized as a "Shitpost" or "Meme." Other users that are more

interested in serious content can filter by "DD" or "Technicals." Oftentimes these post tags can

be quite helpful for users that might not have the financial or WallStreetBets cultural knowledge.

Take the example of the post in Figure 10 tagged as a shitpost. The joke here is that they are

applying the principles and vocabulary of technical analysis for stocks to the number of

COVID-19 cases. Without the shitpost tag, a user may perceive this as a genuine time series

analysis to suggest that the COVID-19 spread will accelerate potentially impacting the market

further, but really they are joking that they are trading shares in the virus.



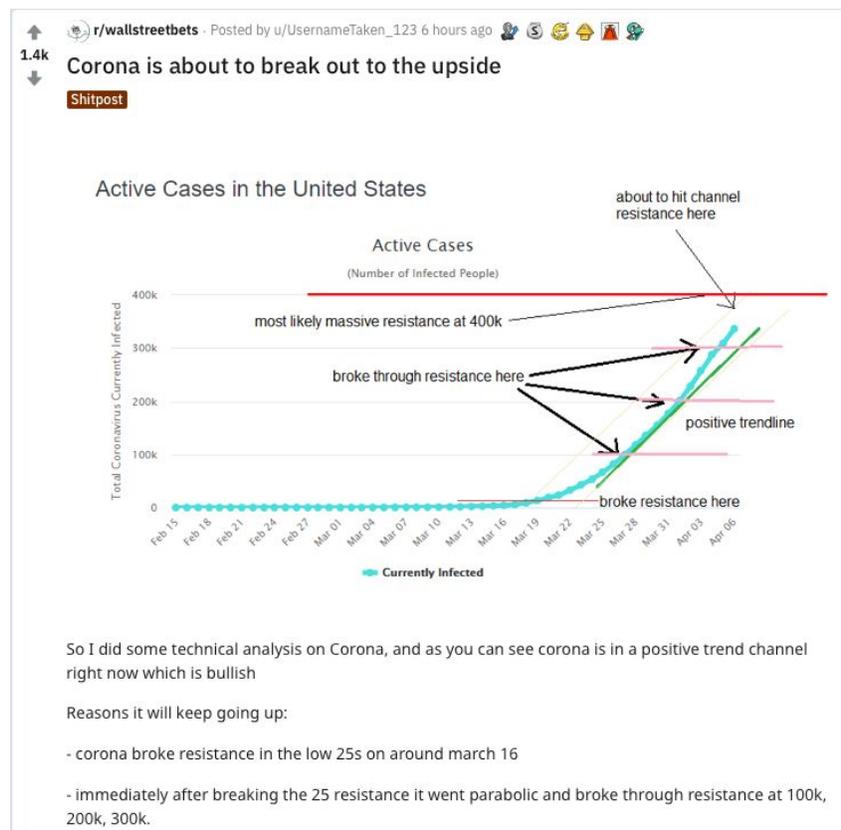

Figure 10: A Sophisticated Coronavirus Shitpost

## 3. Profiles

**Barriers to entry.** Given the position of WallStreetBets as a subreddit in the broader sphere of Reddit as a whole, there is no such thing as a WallStreetBets "profile" only the user's Reddit profile. As such, no new information is collected in order for the user to become a part of the community and they are free to comment and post without explicitly joining WallStreetBets. Even if users are not members of Reddit, they are free to browse everything in the subreddit, but are simply restricted from posting or commenting as they would be in any other subreddit. As such there are virtually no explicit barriers to entry, but there are implicit barriers to entry covered earlier. Users are expected to understand the jokes and linguistic style of community



before posting. If they do not adhere to these standards, they are often met with comments that just simply say "ban" and sometimes an actual ban. Users who receive this punishment are free to create another profile and rejoin the community. The founder even says in terms of those who were banned "[We] don't try to actively hunt them down. We're fine if they create another account with a number two at the end and they continue posting, whatever."

**History.** Since the "profile" of WallStreetBets users is simply their Reddit profile, all profile history is captured in their post and comment history where one can see all of their activity on subreddits in which they participate. This history captures all of their comments, posts, and awards they have received. The only thing in the profile that explicitly quantifies participation in the subreddit is community awards received that are unique to WallStreetBets such as the "golden fuckboy" award.

## 4. Code of conduct

WallStreetBets features a set of rules for the subreddit overall as well as specific guidelines for best practices on what is acceptable content for a post and how you should choose appropriate flair for your post. On the surface, these expectations seem trivial and straightforward, however, there is room for interpretation and the rules are often not enforced uniformly throughout. More specifically, "the moderators don't care if it seems unfair," (Pater et al., 2014) much like in the case of SomethingAwful. This observation in particular relates to Jaime's description of how WSB moderators enforce certain rules. He mocks Reddit users who complain on other subreddits because of their "freedom of speech" or "the First Amendment." Instead, he laughs and explains that the attitude in the WSB moderation team is that of "no, dude,



your question was stupid. There is such a thing as a stupid question and that was stupid. You're not welcome here."

Most of the participants are not particularly familiar with the rules of the subreddit. For instance, Nathan-T1 said he is "vaguely" familiar with the rules because "[the moderators] tend to let pretty much anything fly" and he's never felt inclined to pay attention to them. Similarly, Allen does not know rules, but assumes he is "not a cancer to the sub" because he has not gotten into trouble.

**Constitution.** The r/wallstreetbets sidebar features the following rules, as of April 2020:

1. No Market Manipulation

2. No Pump & Dump, Crypto Discussions, Schemes or Scams

3. No Bullshitting

4. Don't Glorify Losses

5. No Self-Promotion, Social Begging

6. Bad Positions Screenshot

7. Submission Guideline

8. No Generic Memes, No Preschool Memes

9. Political Bullshit

When we spoke to Jaime, he bluntly described the most important rules of the sub to be "no market manipulation, no penny stocks, no pump-and-dump, [and] no scams." He also acknowledged that the rules "evolve over time based on tendencies." For instance, Jaime explained that "some people made the argument that [the current size of the subreddit] could have a measurable impact on the market," and because of this the moderation team had to "add



the necessary measures" to ensure that there would be no "funny business" happening in WallStreetBets. On the other hand, Bacon believes that the most important rule of the community is to "first and foremost be open minded," as well as to "prove your assertion." Bacon claims that the community expects users to be genuine because "[they] don't want to see people paper trading, or [they] don't want to see people doing CSS exploits that will show their accounts being higher than they are." Bacon further emphasized the fact that the community does not tolerate phonies by explaining that "WallStreetBets is not a place to expect that you're going to walk out unscathed if you're stupid. Or if you try and scam somebody." Regulars at WSB agree with this sentiment, as was discussed earlier through the culture of "proof or ban." Furthermore, Max argues that nobody cares about paper trading results because "it's like comparing real soccer to someone's FIFA Xbox recording."

Bacon and haupt91 both suggest that the community itself does a good job at keeping each other in check. Haupt describes WSB as "a collaborative effort that keeps some basic guidelines in motion," and argues that "the rest should be made up by the community, and what they find funny, and what organically comes from that community." He explains that there should never be "a dictator of WallStreetBets" because "what brings people there in the first place is beyond the grasp of any one single person." Bacon explains that the community does not take kindly to jerks, "but if you're going to be a dick, at least be funny." During our participant observation, we noticed that WSB users were quick to denounce and flag any particular posts they did not find funny, which the moderators would then remove. For instance, we saw that a post with racial slurs was met with hostile comments, which invited the original poster (OP) to "sell [their] calls and invest in education" because "racism will not be tolerated." Figure 11



shows another response to this racist thread in which a user explained to the OP why the community did not like what they said. This user suggested that, although they call each other offensive words, they do not use these words to describe, for example, real people with a certain condition.

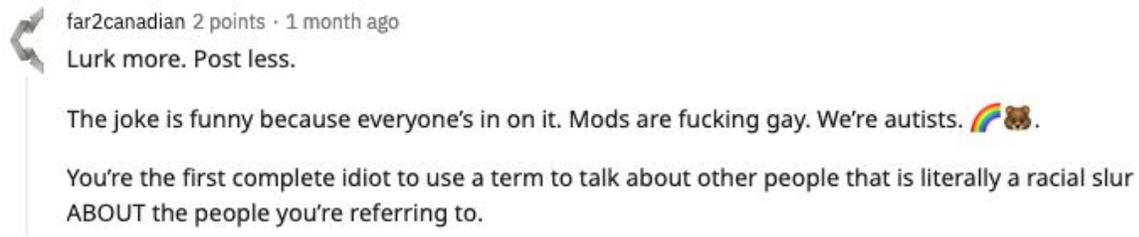

Figure 11: WSB user expresses their dislike for a racist post with the community-approved vocabulary.

Overall, WSB users view the moderation team positively. The moderators are addressed in homophobic terms, and "gay" is equally used as an insult as well as a term of endearment. This practice takes the public humiliation approach as seen in SomethingAwful to a new level. For instance, Pater et al. discuss that the writing of several rules in SomethingAwful "find ways to both humiliate based on actions taken by users as well as attack potential attributes of those users" (2014). In the case of WSB, we have seen many examples of how the community humiliates users based on their actions, and more so based on potential attributes (e.g. "autist"). However, it appears to be particularly unique of this community to openly humiliate the moderators in such offensive (and paradoxically endearing) terms.

Haupt explains that moderation in WSB is "tough" because "on one hand [they're] trying to appease and appeal" to a demographic of "people that like edgy humor," but they must also "be pretty strict about certain things in order to avoid compliance issues or serious ramifications



from Reddit's perspective." Many users believe that the moderators do a good job of balancing these two interests. Max described the moderation as "protective in a good way" because they "are writing tensorflow code to filter posts automatically to save the community from bullshit posts" as well as "actively trying to get rid of people causing harm, or posting fake [content]." During one of the many trolling stunts by the moderators, they disabled the AutoModerator, which resulted in vast amounts of low quality posts being published on the subreddit. Because of this, users begged the moderators to take action, and deeply lamented the absence of the AutoModerator's filtering capabilities.

## 5. Cyclic events

At WallStreetBets, Friday is the community's least favorite day of the week and Monday is the day they look forward to the most (see figure 12 for an example). The nature of the stock market, along with the subreddit's loyal following, make the daily events in the community highly successful. Each weekday the AutoModerator will publish a *Daily Discussion Thread* when the market opens and a *What Are Your Moves Tomorrow* post at market close. According to IS_JOKE_COMRADE these daily discussion threads are "the center" of the subreddit's content. Even the all time top posts in the community rarely garner as many comments as these daily discussions. Although the growing popularity in the sub has resulted in a plethora of unrelated comments overwhelming the daily discussions, members typically use these threads to discuss the market conditions and their trading strategies. Much like the rest of WSB, "locker room" banter saturates the comments in these conversations.



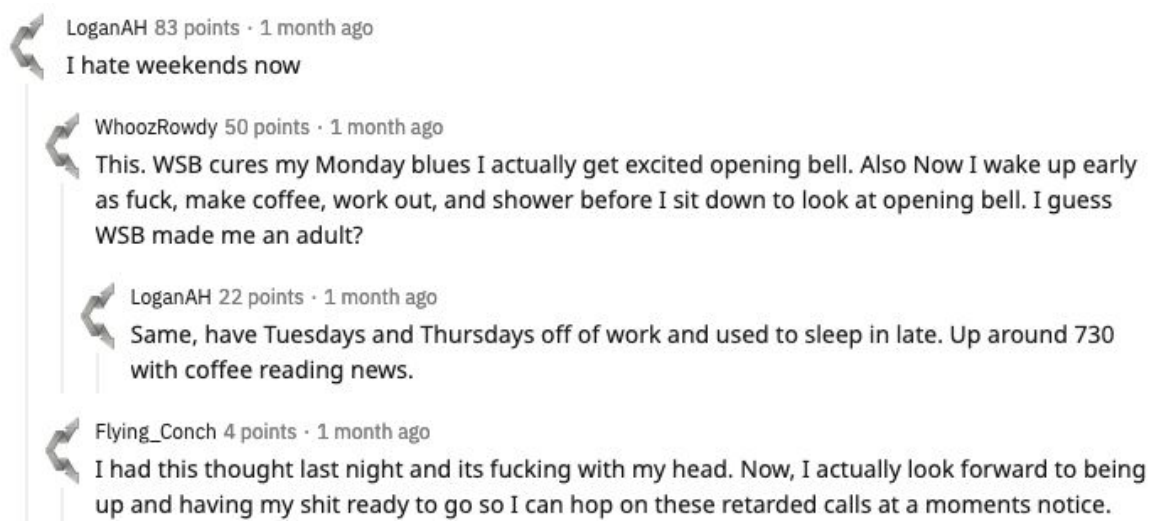

Figure 12: Comments in a WSB thread titled "How has r/wallstreetbets changed your life?"

One of the most active moderators in the community, stormwillpass, is in charge of posting the *Weekend Discussion Thread* every Friday afternoon around market close. Sometime in early 2019, this moderator began the tradition of awarding the first three commenters of each weekend thread either their first user flair in the community, a change of their existing flair, or the ability to designate a flair for another user. The moderator team will also often use the weekend thread to highlight general community announcements as well as encouraging messages. Much of the comments in the daily and weekend threads feature self-deprecating humor where the users seek support for trades-gone-wrong and losing money. In particular, because of the unprecedented volatility in the market during these times, the weekend threads now include numbers and links to resources such as the National Suicide Prevention Lifeline, the National Problem Gambling Helpline, and the Substance Abuse and Mental Health Services Administration's National Helpline.



The mob mentality of the subreddit also facilitates the development of temporary and sporadic daily or weekly events. For instance, on Monday March 2, 2020, Robinhood–the most popular broker in the community–experienced a platform-wide outage[9]. This interruption prevented users from accessing their trades for the entire day, which (for many reasons beyond the scope of this paper) resulted in a lot of WSB users losing money–although a few benefitted from keeping their options trades for longer than they initially intended. The Robinhood support team emailed their customers explaining that the situation had been resolved, and subsequently experienced another crash when the market opened the next day. There are several design features, which Rogozinski covers in his book, which make Robinhood the butt of plenty of jokes on WSB. However, the amount of comments targeted at Robinhood on the week of their outages prompted the subreddit's moderators to create corresponding *Robinhood Crash Megathread* posts for each day of the outage. Because WSB users doubted Robinhood's ability to recover from these interruptions, the moderators also created several daily posts throughout that week with the cheeky title of *Preemptive Robinhood Crash Thread*.

**Contests.** WSB moderators often host charity efforts through WallStreetBets. Typically, they ask for users to share valid proof of their contributions to charity, and a moderator will compensate them in the form of a flair. Recently, one of the moderators hosted a charity livestream, for which the end goal was to raise $69,420 (a reference to the community's favorite numbers) for the organization Autism Speaks. In the end, they raised over $33,000 after hosting the livestream for over nine hours. The moderator who planned the event (WallStreetBooyah) described the event as contributing to a "tip jar" for all the times they misused the word "autist" in the community.

---

[9] https://blog.robinhood.com/news/2020/3/3/an-update-from-robinhoods-founders

Ever since they gained mainstream popularity, WSB started the tradition of a faux-competition where the regulars (traditional users and moderators alike) will trick newbies into a "trap," and subsequently ban them if they actually fall for it. This is a community effort to reduce the amount of newcomers who do not lurk enough to understand or (ironically) respect the local culture. This resembles the approach in SomethingAwful, where "new members are also more susceptible to violating community standards and getting probated or banned" (Pater et al., 2014). Haupt was reluctant to explain this "anti contest" phenomenon in WSB, but stated that "[they] have like a bloodbond almost. [They] just look at each other and nod. And it's always thought, but never spoken why [they] conduct this annual sacrifice."

## 6. Roles

Due to the nature of Reddit as a platform, it is not helpful to discuss the various roles of WSB members as it relates exclusively to the permissions or the type of access they have to the site–anyone and everyone can view any post or comment on the subreddit. We will instead focus this discussion on roles to expand on the actions each user can perform, and the affordances on said actions granted by the structure of the community, as well as the nature of the status and the representation of each role in WSB.

In her book, Kim describes the importance of the roles in online communities. She claims that "communities are held together by a web of social roles, and you can help your community flourish by providing features and programs that support these roles" (Kim, 2000). Kim outlines



five roles which exist in all online communities: visitors, novices, regulars, leaders, and elders.

From the dynamics we observed on WallStreetBets, we found it helpful to combine these roles

into two overarching categories in order to simplify our analysis. These are explored in the

remainder of this section.

**Visitors and regular users.** Kim describes visitors as people who arrived at the

community with different needs, and have arrived in different ways such as a web search or

friend's recommendation. In the case of WSB, visitors are often Reddit users who found an

outrageous post through r/all or perhaps journalists and even finance professionals. Visitors

quickly learn that WSB is "not for people with thin skin" (Frank). Other interviewees described

their first encounters with the subreddit as "pretty childish, crude humor, and surprising honesty

in people's profit and losses" (Allen), and described their first impression of its users as

"degenerate gamblers" (IS_JOKE_COMRADE). Perhaps the most accurate summary of the first

impression WSB gives to a novice user was given by Nathan-T1:

> "[I] was pretty confused what it was all about when I first visited it. Sub kind of has its
>
> own language and cult like following so it was kind of hard to understand what people
>
> were saying at times. Definitely did not want to ask any questions or you would get
>
> roasted"

Nathan-T1 realised that WSB is not the place to post meaningless questions, and the best strategy

to become used to the culture is by observing the regular users.

Like with other subreddits, WSB visitors do not need any special permissions to browse

the content of the subreddit. They can participate by reading the subreddit and by looking for

advice. However, Reddit does not permit users without an account to post content on the



platform. Therefore, in order to participate visitors need a Reddit account. Visitors who have a

Reddit account and decide that they like the content on the subreddit can join WSB. One of the

benefits of joining a subreddit is convenience. Frank joined WSB after several visits because the

subreddit "automatically shows up on [his] homepage."

Kim suggests that novices should be educated about the community life and protected

from getting into trouble. However, this is not the case in WSB. The community's culture of

trolling does not easily forgive mistakes made by novice users. One of the participants, Max,

cautioned that when you first visit WSB and begin participating, you should be aware of the

"ubiquitous sarcasm" because you may not realize that someone is joking and may take their

words "as serious advice or a tip on something". Other interviewees also advised newcomers to

be careful when they first start trading options and participate in WSB. Frank's advice to the

newcomer is "don't ask any basic, technical questions because a lot of the people that you see

posting these things don't know about the fundamentals and really how easily you can lose

money". IS_JOKE_COMRADE suggests that newcomers should become familiar with trading

which they could to by "sign[ing] up for a paper trading account. Fake trades. Read the posts,

understand options. It'll happen organically."

Kim describes the regular users as "established citizens." The regular users we

interviewed interact with the subreddit often. Some lurk most of the time, while others post and

comment throughout the day. Distinguished regular users are easy to identify by their community

flair. One of the participants described the flair as "some kind of recognition" (Henry). He

claimed that "no matter what the flair is," WSB users will give you more attention because you

seem "probably more experienced than [the] other sheeps" when you have a flair to your name.



**Moderators and administrators.** The lighthearted nature of WSB–that is not taking anything or anyone seriously–is evident across all strata of the community. The moderation team is no exception. For instance, there have been cases where individual moderators have started threads that requested users to comment, and would grant moderator permissions to anyone who participated in said thread. Other instances of instability for the sake of a joke have happened such as when Jaime "had moments where [he] just fired all the mods and [then] just rehired them all." According to Kim "not all regular [users] will want to take the leadership position, but those who do will relish the status and visibility that comes with an official sanctioned role." Despite this foolish yet deliberate chaos, the design of the reddit platform as a whole allows users to view the current list of moderators as well as their specific permissions on the subreddit.

Aside from posing what is probably the most ridiculous flair in the community, the WSB moderation team also possesses a unique character, AutoModerator (or AutoMod). Although this account is not unique to the subreddit, the WSB moderators customized everything they could. For instance, AutoMod has displayed flairs in the same flavor as the rest of the culture of the subreddit, such as "Destroyer of Worlds" and "Retard Wrangler." AutoMod is also configured with automatic responses, such as "Eat my dongus, you fucking nerd!" whenever a user comments the keyword "sticky" (or its derivatives) in a WSB thread. WSB users will often engage with AutoMod in the comments, and even go as far as awarding him with gold on the daily discussion threads.



## 7. Real world integration

When WallStreetBets began, Jaime participated in a chatroom along with fellow WSB members, where he claims to have spent in total hours "just as much time [chatting] with them as [he] had with [his] best friends that [he] knew since [he] was a child." Eventually, most of the members from this chatroom became the first cohort of moderators in WSB. Jaime explained that a lot of them gathered at various meetups around the US and Mexico, and even a couple of them attended Jaime's wedding in Mexico.

As was mentioned in the cyclic events section, the WSB moderators will often organize contests, which become new or different creative avenues to share more finance-themed jokes. A notable event that integrated with members' real world was the WSB limerick contest, which resulted in a moderator mailing profane gifts to the winners. The physical traits of these gifts literally embodied the irreverent images that were most popular within WSB at the time. Bacon recounted that the WSB moderators have "all had [their] spin on stuff, [and] that was one of [his] contributions." However, Bacon acknowledges that some of these events and prizes, which the community once celebrated, cannot happen again because they are no longer "with the times," and "not something that [will] generally be viewed as a joke anymore." He recognizes that this is due to the popularity of the sub, which often reaches among the top three in total Reddit traffic, as well as their growing presence in the "the serious, legit financial community."

The latest way through which the subreddit is permeating the real world is through the (infamous in the community) WSB championship[10]. The sponsors advertise the championship as an esports-like high-risk trading competition, which will consist of twelve contestants and one

---

[10] https://wallstreetbets.net/



"large cash prize" for the winner. The event promoters themselves are also infiltrating the community in the form of ads and discount codes for True Trading Group[11], a company which claims to help users "beat the market" by sharing weekly trading strategies for a premium.

## 8. Leadership

Much has been said about the leadership in WallStreetBets. To summarize, a team of moderators work disjointly with no codified strategy. They all are left to make moderation decisions as they see fit. According to Bacon, Jaime and the senior moderators mostly spend their time on Discord and IRC hanging out and thinking of "cool stuff" that "advances the community" leaving the junior moderators to deal with the everyday moderation tasks. During these conversations the mods plan events and changes to the community.

Bacon describes Jaime as having a major role as the community founder since he has some "skin in the game" as his real name is forever linked to the group. As such, he has often been responsible for making important announcements and screen sharing with users to verify that they have not faked their gains. Jaime admits that much of what we see today on WallStreetBets has little to do with him. He left the community several years ago when it was significantly smaller only to come back within the last year leading to one of our interviewees to call him "the deadbeat dad that came back." Jaime claims he came back because the community needed "guidance" and he sees his responsibility as providing this aforementioned guidance.

There have been recent disputes about the leadership of the subreddit. As was mentioned earlier, Jaime was responsible for organizing the WSB championship competition sponsored by True Trading Group. He planned to sell tickets and pay-per-views. Jaime additionally used the sidebar of the subreddit, which has traditionally been ad free, to advertise his book, the competition, and the True Trading Group. Many users were upset that Jaime appeared to be using the subreddit for personal gain and not being transparent about where the money from these endeavors was going. One of the most important rules of the group was "No schemes or scams," yet it appeared to many users as though the founder was breaking these rules.

---

[11] https://truetradinggroup.com/

Bacon reached out to us after our interview to inform us of the backstory behind what Jaime was doing. He claims that Jaime made these decisions unilaterally against the wishes of all the senior moderators as the sole administrator of the subreddit. Additionally, he says that Jaime added members of the True Trading Group as moderators giving them the ability to remove posts critical of the competition or True Trading Group in general. After some push back from the senior moderators, Jaime reportedly took away the moderating permissions of all of the senior moderators. As of writing this, it appears that the senior moderators have been able to get the Reddit administrators to remove Jaime as a moderator in the subreddit and get their moderator permissions reinstated. They are currently in the process of redesigning the subreddit leaving the message seen in figure 13 and taking the subreddit temporarily private.



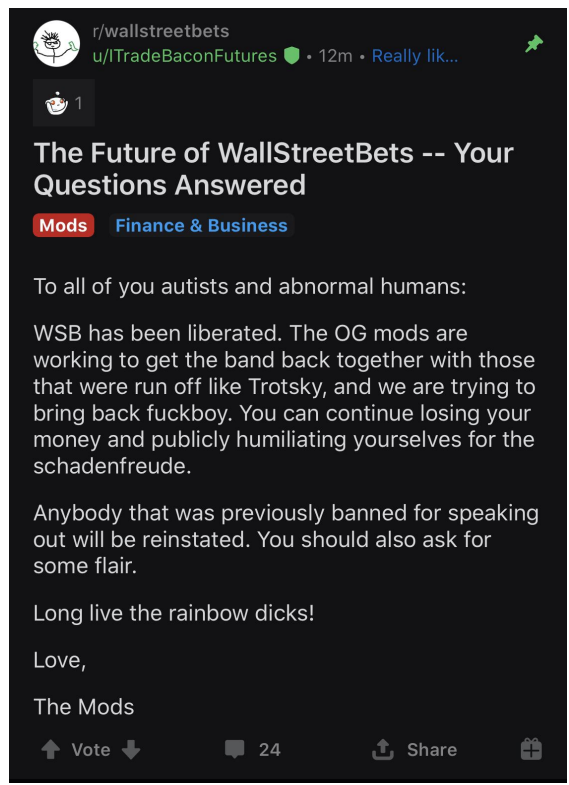

Figure 13: Bacon's Statement on Jaime's Removal

This power struggle says a lot about the leadership of the subreddit. This was never designed to be a place run by a central authority. Before any of this had even happened, haupt91 said "I don't think there ever should be like a dictator of WallStreetBets" and that it should instead be "a collaborative effort that keeps some basic guidelines in motion and then the rest should be made up by the community." Oldenburg states that third places create a sense of "at-homeness" that often results in those who claim a third place feeling "a sense of possession and control over a setting that need not entail actual ownership." Jaime's actions appear to have threatened the ability of many users to feel at home here, as many felt they no longer had any semblance of control over the direction of the subreddit. The actions of the senior moderators to



push Reddit to remove Jaime and their subsequent announcement (figure 13) display their belief that this community belongs to all the members.

## Conclusion

WallStreetBets seems like a community that on paper should not be successful. It focuses on a very specific, high-risk style of trading that is essentially gambling. It is populated by users who constantly flame and insult each other with slurs. It actively purges new members and takes steps to make them feel unwelcome. It offers no transparency in moderation decisions. It empowers those pushing the boundaries of acceptable behavior. Yet, nearly 1.1 million users found a reason to join and stay over the years.

It's undeniable that WallStreetBets means a lot to it's users. It's not a place to come, gather some trading ideas, and leave. It's a third place for people–mostly young men–who want to bring some excitement and edge into their otherwise normal, professional lives. WallStreetBets provides an outlet where they can, as one user perfectly put it, "be an asshole with some friends." The unique, comical fixation on the ever changing market pushes the community to constantly create new and relevant content leading to an engaging and rapidly evolving humor landscape. This creates a place that is always good for a quick laugh in good company.